\documentclass[12pt]{article}
\usepackage{amsmath, amssymb}
\usepackage[margin=2.5cm]{geometry}
\usepackage{graphicx}
\usepackage{authblk}          
\usepackage[hidelinks]{hyperref}
\usepackage{natbib}           
\title{Incoherent Diffraction Imaging with a Pseudo-Thermal Light Source: supplemental document}
\author[1,2]{Pablo San Miguel Claveria$^{*}$}
\author[3]{Peer Biesterfeld}
\author[1]{Sebasti\~ao Antunes}
\author[1]{Matilde Fernandes}
\author[1]{Matilde Garcia}
\author[1]{Matilde Nunes}
\author[1,2]{Lucas Ansia Fernandez}
\author[1]{Gareth O. Williams}
\author[3]{Sven Froehlich}
\author[4]{David Theidel}
\author[3]{Philip Mosel}
\author[5]{Ihsan Fsaifes}
\author[3,6]{Andrea Trabattoni}
\author[7,8]{Marco Piccardo}
\author[5]{Jean-Christophe Chanteloup}
\author[3]{Milutin Kovacev}
\author[4]{Hamed Merdji}
\author[1]{Marta Fajardo}

\affil[1]{GoLP/Instituto de Plasmas e Fus\~ao Nuclear, Instituto Superior
  T\'ecnico, Universidade de Lisboa, Lisbon 1049-001, Portugal}
\affil[2]{Instituto de Fusi\'on Nuclear ``Guillermo Velarde'', Universidad
  Polit\'ecnica de Madrid, Jos\'e Guti\'errez Abascal 2, 28006 Spain}
\affil[3]{Institut f\"ur Quantenoptik, Leibniz Universit\"at Hannover,
  Hannover 30167, Germany}
\affil[4]{Laboratoire d'Optique Appliqu\'ee, ENSTA Paris, CNRS, \'Ecole
  Polytechnique, Institut Polytechnique de Paris, Palaiseau, France}
\affil[5]{LULI, CNRS, \'Ecole Polytechnique, CEA, Sorbonne Universit\'e,
  Institut Polytechnique de Paris, 91120 Palaiseau, France}
\affil[6]{Center for Free-Electron Laser Science CFEL, DESY, Hamburg, Germany}
\affil[7]{Department of Physics, Instituto Superior T\'ecnico, Universidade de
  Lisboa, Lisbon, Portugal}
\affil[8]{INESC MN, Lisbon, Portugal}

\date{}   

\begin{document}

\maketitle

\begin{abstract}
Incoherent Diffraction Imaging - IDI - is a lensless imaging technique that has been recently proposed to exploit the partial coherence of incoherently scattered light to retrieve structural information from the scattering centers. Similar to the stellar intensity interferometry of Hanbury Brown and Twiss, the signal builds up on the second-order spatial correlations of the emitted light. The complex spatial distribution of the target is thereby encoded in the spatial intensity fluctuations of the scattered light. The first experimental realisations of this imaging technique have been realised using the fluorescence excited by an ultra-short X-ray pulse at Free Electron Laser (FEL) facilities. Here, we propose an alternative setup based on a table-top Pseudo-Thermal Light Source that works as an optical emulator of the FEL experiment. This setup allows us to explore IDI under a wide range of physically relevant conditions as well as to benchmark numerical and analytical models currently used to determine the imaging capabilities of this technique.
\end{abstract}

\section{Introduction}
The coherence of light, or the spatial and temporal phase carried by a light source, has been decisive for imaging at short wavelengths, where diffraction-limited optics are unavailable.
Coherent diffraction imaging (CDI), which makes use of the diffraction pattern of coherent light from an object to retrieve the real space image of the object, relies on sources with high degree of spatial coherence, reaching spatial resolutions limited only by the wavelength.
After its first experimental demonstration \cite{Miao_Nat_1999}, CDI has successfully validated single shot femtosecond nanoscale imaging~\cite{Chapman_NatPhys_2006,Ravasio_PRL_2009} with the possibility to extend to 3D imaging~\cite{osti_807154, Robinson2001, Duarte_nat_phot}. Especially, illumination with short wavelengths like X-Rays for imaging of biological samples has been widely adopted~\cite{Miao2012,Nishino2009}. 
 Storage Ring (SR) sources have played a central role in the experimental development of CDI, particularly in scanning and Bragg scattering modes. Algorithmic developments have further shown that limited partial temporal and spatial coherence can be accommodated within CDI frameworks~\cite{Abbey_NatPhot_2002}.

Recently, a new lensless imaging technique has been proposed~\cite{Classen_PRL_2017} to exploit second-order (intensity) correlations of partially coherent light. In analogy to CDI, it was named Incoherent Diffraction Imaging (IDI) to reflect the fact that the photons are incoherently scattered, while still allowing imaging under conditions of partial coherence. This technique requires the recording of the scattered light within a timescale shorter or comparable to the coherence time, $\tau_c$, of the light field. Hence the recorded diffraction pattern can be considered stationary during the acquisition time, as naturally happens in CDI. Whereas the diffraction pattern will fluctuate for each individual acquisition, the autocorrelation of the recorded intensity is insensitive to these spatial variations and will thus progressively build-up when averaging over multiple acquisitions. Being based on intensity correlations, this technique shares the working principle with the intensity interferometer originally proposed by Hanbury Brown and Twiss to overcome atmospheric distortions on the light coming from the stars~\cite{Brown_1968,Twiss_1969}. However, instead of using two spatially separated Photo-Multipliers Tubes, the spatial correlation is carried out on the intensity of the light measured by a pixelized detector.

One of the most promising applications of IDI is to retrieve the structural information of an object via incoherent scattering processes, namely fluorescence~\cite{Ho_Struct_2021,Wollweber_SD_2024}. Fluorescence is typically considered as incoherent light due to its quantum nature, its emission originating from a collection of independent (uncorrelated) sources. The coherence time of the fluorescence emission is given by the radiative lifetime of the corresponding transition, i.e. inversely proportional to the spectral linewidth $\Gamma$ ($\tau_c = 2\hbar/\Gamma$)~\cite{white1934introduction}. For inner-shell fluorescence of transition metal atoms - one of the most narrow linewidth emissions- one finds that $\tau_c$ is of the order of fs (0.82 fs for Fe, 0.62 fs for Cu). Most advanced FELs facilities are currently able to produce X-ray pulses with pulse durations of the order of these timescales, and thus the fluorescence excited by these FEL pulses can be temporally gated for durations of the order of its coherence time. Therefore, in a single shot, a good degree of temporal coherence can be obtained from inner-shell fluorescence in solids excited by a fs FEL pulse. However, each short fluorescence event will produce a different relative phase between positions on the emitted wavefront, causing variations in the pattern on the detector from shot to shot. Contrary to CDI, the accumulation of multiple patterns on the detector does not result in an improved signal to noise ratio, rather a loss of information as the stochastic spatial variations in each shot lead to blurring of any scattered signals when accumulated.

In order to increase the visibility of the measured spatial pattern, IDI relies on accumulating the second-order spatial correlations of the light field, i.e. the spatial auto-correlations of the recorded intensity, which are insensitive to the phase variations between each fluorescence event~\cite{Trost_2020}. In this way, the structural information of the emitters of incoherently scattered light can be retrieved from the second order correlation by recalling the Siegert relation~\cite{goodman1985statistical} for a Thermal Light Source (TLS) within the limit of a large number of emitters~\cite{Trost_2020}, that relates the first and second order correlations as

\begin{equation}
    g^{(2)}(\vec{q})=1+\beta | g^{(1)}(\vec{q})|^2
    \label{eq:Sieg}
\end{equation}
where $g^{(2)}(\vec{q})$ is the normalised intensity auto-correlation and $g^{(1)}(\vec{q})$ is the normalised scattering amplitude (the strength of the coherent diffraction pattern at a particular object's spatial frequency $\vec{q}$). The visibility parameter $\beta$ accounts for the partial temporal coherence of the recorded light ($0<\beta\leq 1$) and for the degree of polarization. For a fully coherent illumination with a single polarization state, the visibility $\beta=1$.

Other works have used intensity correlations from inner-shell fluorescence of Cu atoms excited by a short FEL pulse to determine the transverse focal spot size~\cite{Nakamura:wz5010} and pulse duration~\cite{Inoue:gb5094} of the incoming X-ray pulse. The potential of IDI for 3D reconstruction of the intensity of an FEL pulse along the focal region has also been recently demonstrated at the CXI station of the LCLS FEL~\cite{radloff_2025_arxiv}. Moreover, the first imaging experiment of a non-trivial structure using IDI~\cite{Trost_PhysRevLett.130.173201} was carried out at the MID instrument of the European XFEL~\cite{Madsen:ay5570}. By focusing the FEL pulse onto two spatially separated focal spots on a thin Cu target, the associated second-order spatial fringes were observed in the intensity correlation map after accumulating $5.8\times10^7$ fluorescence frames. From this $g^{(2)}(\vec{q})$ map, the size and separation of the two X-ray spots was correctly reconstructed. 

The large number of frames and the associated long acquisition time ($\approx5$h) are a consequence of the low number of averaged detected photons per pixel per frame $\mu=7.7\times 10^{-3}$ and the rather long FEL pulse duration (6.2 fs, compared to the Cu fluorescence $\tau_c\approx$ 0.6 fs), which in this experiment ultimately resulted in $\beta=0.018$. According to analytical and numerical models~\cite{Trost_2020}, these long acquisition times can be greatly reduced with shorter excitation pulses and with brighter fluorescence sources (3 minutes for 3 fs pulses~\cite{Classen_PRL_2017}). These long acquisition times, coupled with the high demand for XFEL access, means that imaging of more complex objects has been so far elusive. Furthermore, rigorous tests of the imaging performance of IDI for physically relevant parameters, such as coherence time and source brightness, are needed to fully understand the imaging capabilities of this new technique. In this letter, we demonstrate a table-top experimental setup based on a Pseudo-Thermal Light Source to experimentally explore IDI and benchmark analytical and numerical models. Although Pseudo-Thermal Light Sources have been explored for diffraction-based imaging~\cite{Katz_NatPhot_2014,Schneider_Nat_phys, Lu_OptRev_2021}, here we extend the use the PTLS to experimentally study the effect of the second-order temporal coherence on the visibility $\beta$ of the retrieved $g^{(2)}$, which can then be extrapolated to the short-wavelength regime where diffraction-based imaging techniques are of most use. The first section describes the proposed optical setup, including a thorough characterisation of the degree of coherence of the PTLS. The second section presents acquisitions that demonstrate IDI's capability to encode structural information in the second-order correlation maps $g^{(2)}(\vec{q})$. 
Finally, we provide quantitative measurements of the imaging performance of our IDI setup over a wide range of physical parameters, comparing these measurements with existing analytical and numerical models. Following the theoretical framework of~\cite{Trost_2020}, we assess imaging performance through the Signal-to-Background Ratio (SBR) and Signal-to-Noise Ratio (SNR) of the retrieved $g^{(2)}$ maps, which are the relevant figures of merit for IDI at this stage of its development. In this way, we aim to provide a new experimental approach, as well as experimental evidence of the feasibility of the method.

\begin{figure}[h]
    \centering
    \includegraphics[width=0.99\textwidth]{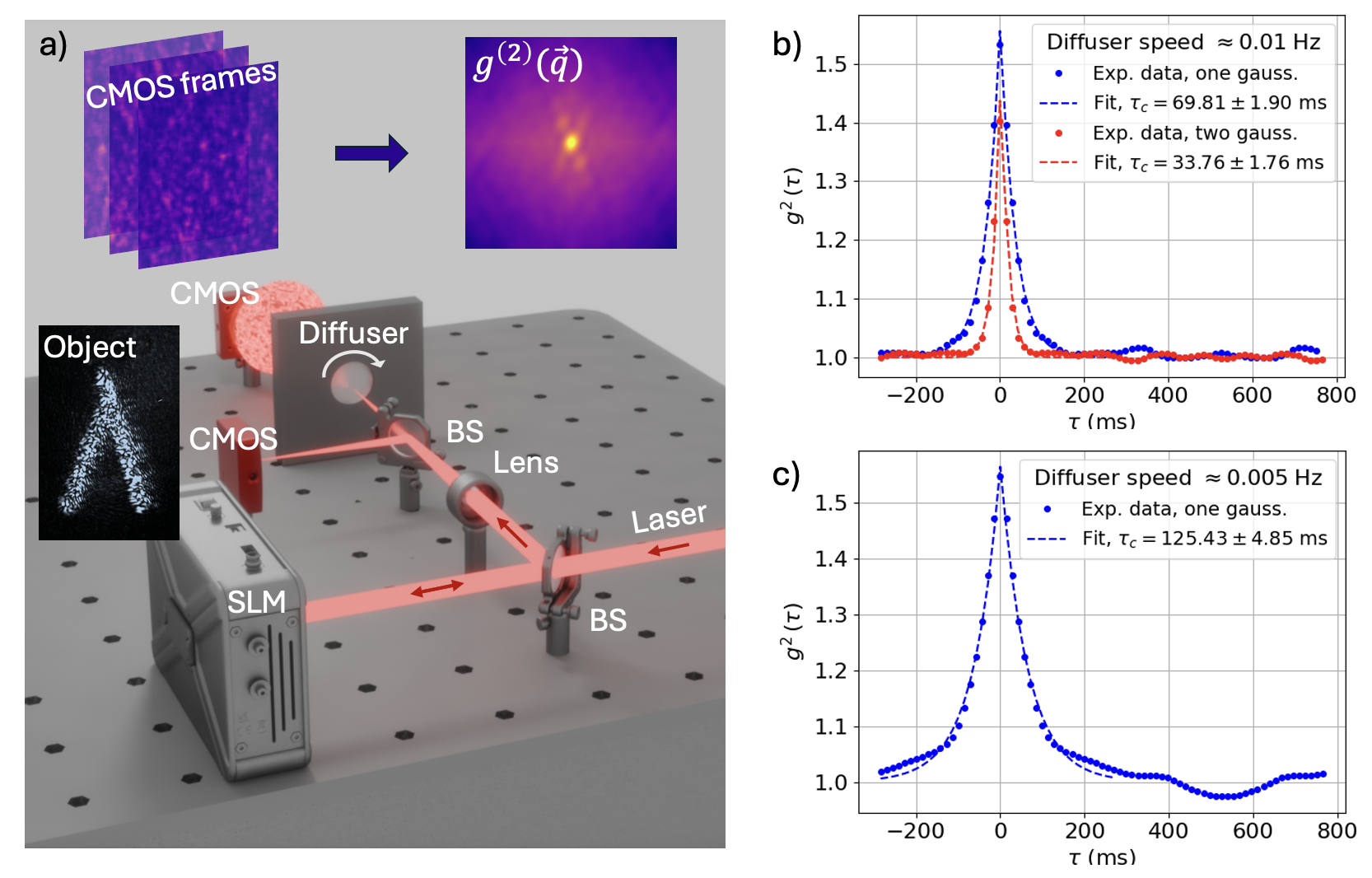}
    \caption{(a) Sketch of experimental setup. The laser comes from the right of the image, goes through the beam splitter (BS), reflects at the SLM chip (where it acquires the holographic phase corresponding to desired object intensity), reflects from the same BS towards a plano-convex lens and is then focused on the rotating diffuser. An additional BS is placed between the lens and the diffuser to monitor the laser intensity at the focal plane of the lens. Two CMOS cameras are used to record the aforementioned laser intensity at the focal plane of the lens as well as the diffused light. The inset show examples of object intensities, three frames of recorded diffused light, and the $g^{(2)}$ auto-correlation map computed from 100 frames. (b) $g^{(2)}(\vec{q}=0,\tau)$ experimental measurements for a single and double Gaussian intensity distributions impinging onto the diffuser, and associated Lorentzian fits from which the $\tau^{(2)}_c$ is extracted (see legend). (c) Same as (b) but for a diffuser rotating at twice the speed of (b).}
    \label{fig:fig1}
\end{figure}

\section{Experimental setup}

In order to build a Pseudo-Thermal Light Source with a large degree of tunability in terms of coherence, we have used a ground-glass diffuser with 1500 grit/mm mounted on a motorised rotation stage, which is illuminated by a linearly polarised CW diode laser of $\lambda=660\;\rm nm$. On the one hand, such a PTLS allows to control the spatial coherence of the diffused light, i.e. the speckle size, via the illumination area (laser spot size) on the diffuser and the distance between the diffuser and the spatially resolved detector~\cite{goodman1985statistical}. On the other hand, for a given illumination area, the second-order temporal coherence can be controlled by the rotation speed of the diffuser. Note that the second-order temporal coherence is also inversely proportional to the illumination area~\cite{Estes_JOSA_1971,Starovoitov_JAS_2023}, and thus for the same diffuser rotation speed different illumination areas lead to different degrees of second-order temporal coherence. Ultimately, the brightness of this PTLS, namely the number of photons per pixel per frame $\mu$ can be controlled via the laser power and additional filtering. 

The diffused light is recorded with a CMOS camera placed at $\approx 18\;\rm cm$ away from the diffuser with a small lateral offset with respect to the optical axis to avoid the specular part of the laser beam that is not scattered. The distance from the pixelised detector to the diffuser was chosen to ensure enough speckle sampling~\cite{Trost_JSR_23}. In this experiment the temporal gating of the incoherent light is carried out via the exposure time of the detector (i.e. we gate the light detection instead of the light emission), which allows us to control the degree of visibility $\beta$ in our IDI acquisitions. In addition, we use a Spatial Light Modulator (SLM) together with a 
lens to holographically control the intensity distribution~\cite{Rosales_SLM} impinging on the diffuser and thus create any arbitrary test object. The SLM-generated intensity distribution that is sent to the diffuser is recorded with an additional camera by picking part of the incoming beam with a beam splitter. A schematic of the experimental setup is shown in Fig.~\ref{fig:fig1}(a). This flexible optical setup allows the retrieval of structural information of the SLM-generated intensity distribution by computing the second-order correlations of the diffused light recorded by the detector.

In contrast to other imaging experiments based on high-order correlations from a Pseudo-Thermal Light Source~\cite{Schneider_Nat_phys}, in this setup the diffuser acts both as a source of pseudo-thermal light and as the object to be reconstructed, mimicking the fluorescence-based IDI experiments. Whereas the use of an SLM to control the intensity distribution sent to the diffuser poses some limitations in terms of minimum object size and resolution, it offers a large degree of versatility to explore different object complexities without the need of machining new object masks. Examples of object distributions created using this setup are shown in Fig.~\ref{fig:fig2}. Note that other devices, such as transmissive LCDs or digital micromirror devices (DMDs), could be similarly used to generate the desired intensity patterns. 

In order to assess the second-order temporal coherence of the source, we measured the second order temporal correlation function
\begin{equation}
    g^{(2)}(\tau)=\frac{\langle I(t)I(t+\tau)\rangle}{\langle I(t)\rangle^2}
\end{equation}
where the average is performed over time $t$ and over the detector pixels. Figure~\ref{fig:fig1}(b) and (c) show the results of these measurements for two rotation speeds 0.01 Hz and 0.005 Hz (blue dots) when a Gaussian intensity pattern impinges the diffuser, as well as the fitted Lorentzian curves whose associated second-order temporal coherence $\tau^{(2)}_c$ is shown in the legend. For these acquisitions, taken with a detector exposure time of 5 ms, we measure a second-order temporal coherence ranging from 10s to 100s of ms. As expected, the retrieved $\tau^{(2)}_c$ is inversely proportional to the rotation speed of the diffuser. Moreover, Fig.~\ref{fig:fig1}(b) also shows the measured second order temporal coherence when two spatially separated Gaussian distributions of the same size as the single Gaussian pattern impinge on the diffuser (red dots). As expected, the retrieved $\tau^{(2)}_c$ is reduced by a factor of two due to the doubled area of the diffuser that is illuminated~\cite{Estes_JOSA_1971}. We would like to point out that a reduced ROI of $100\times 100$ pixels was used in these acquisitions in order to run our CMOS detectors with the required frame rate to resolve the temporal coherence of our PTLS. This limited ROI leads to a reduction of the $g^{(2)}(\tau=0)<2$~\cite{Li_PhysRevA_2020}.

\section{Measuring $g^{(2)}$ with a PTLS}

\begin{figure}[h]
    \centering
    \includegraphics[width=0.99\textwidth]{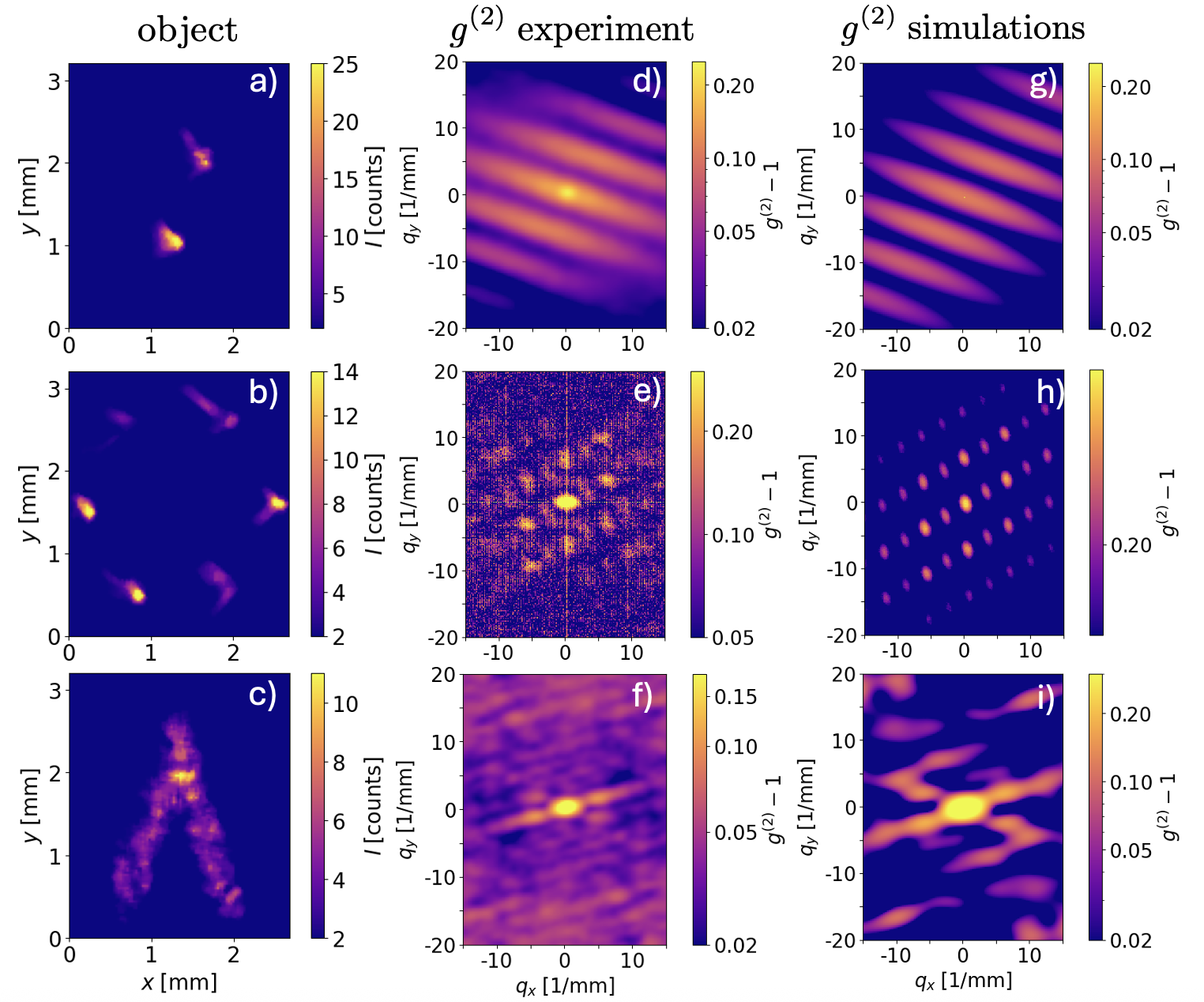}
    \caption{Three exemplary IDI acquisitions performed with our PTLS-based setup. Sub-figures (a,b,c) show the recorded intensity distribution impinging the rotating diffuser. Sub-figures (d,e,f) show the computed $g^{(2)}-1$ from 100 acquisitions of the diffused light. Sub-figures (g,h,i) show the associated computed $g^{(2)}-1$ from 100 simulated frames. See text for further details. }
    \label{fig:fig2}
\end{figure}

We now show how second order spatial correlations can be used to extract structural information of the scatterers distribution given the finite degree of coherence of a PTLS. For these acquisitions, shown in Fig.~\ref{fig:fig2}, we set the coherence time of the PTLS ($\sim 100\;\rm ms$) to be significantly longer than the exposure time of the detector ($ 5 \;\rm ms$) in order to detect a single temporal mode. Moreover, a acquisition deadtime of $\approx 200 \rm \; ms$ is set between frames to ensure that consecutive frames are uncorrelated. In this Fig.~\ref{fig:fig2}, each row corresponds to a different object intensity distribution, which are shown on the first column Fig.~\ref{fig:fig2}(a,b,c).

After computing and averaging the intensity auto-correlations over 10 frames using $400\times 400$ pixels, the normalised $g^{(2)}(\vec{q})$ maps are shown on Fig.~\ref{fig:fig2}(d,e,f). A detailed discussion on the auto-correlation computation can be found in the Supplemental Document. The diffraction patterns associated to each object intensity distribution are clearly visible in these second order spatial correlations, despite the varying level of visibility. It should be noted that for these acquisitions the number of frames is enough to reach the maximum visibility given the large number of photons per pixel per frame $\mu\gtrsim 1$.

We have also performed numerical simulations to assess the level of visibility of the measured second order correlation maps. Similarly to~\cite{Trost_2020}, these simulations use Fresnel propagation to propagate the object intensity distribution measured in the experiments with a randomised phase from the diffuser to the detector plane (see Sec. 4 of the Supplemental Document for a detailed description). In order to account for the degree of temporal coherence, we add the propagated intensity over several randomisations of the initial phase. In this way we can simulate the detection of an integer number of temporal modes $N_{\rm M}$ when the detector detection time is greater than $\tau_c$. Moreover, we include the Poisson statistics associated to the CMOS detector photon detection to generate synthetic frames. This simulation tool is accessible in~\cite{repo_IDI_2024} for researchers to quickly check the feasibility of their imaging conditions.

The $g^{(2)}(\vec{q})$ computed from 10 synthetic frames generated by the corresponding intensity distributions showed in Fig.~\ref{fig:fig2}(a,b,c) with a single temporal mode are displayed on Fig.~\ref{fig:fig2}(g,h,i). Overall these simulated $g^{(2)}(\vec{q})$ maps reproduce well the experimental features. The main source of discrepancy between the experiment and simulations originates from the higher levels of background in the $g^{(2)}(\vec{q})$ maps computed from experimental data, subject to additional sources of background ambient noise, thermal variations affecting dark current levels and readout electronics. We additionally attribute the central bright spot visible in the experimental $g^{(2)}(\vec{q})$ maps [Fig.\ref{fig:fig2}(d,e,f)] to a DC component associated with the finite diffraction efficiency of the SLM and the non-unit scattering efficiency of the GGD, neither of which are included in the simulations. These higher background levels result in the loss of some features of the $g^{(2)}(\vec{q})$, especially as the object structure becomes more complex. Recalling the Siegert relation (Eq.~\ref{eq:Sieg}), the real-space distribution of the object intensity can be recovered using existing phase retrieval algorithms~\cite{Trost_PhysRevLett.130.173201}, or recently developed adaptations for higher-order light correlations~\cite{bojer2023phase,Peard:23}. As a proof of concept, we apply an iterative phase retrieval algorithm (HIO followed by ER refinement, see Supplemental Document Sec. 5) to the experimentally retrieved 
$g^{(2)}(\vec{q})$ maps of the two-dot object. The resulting reconstructions, shown in Fig.~\ref{fig:fig4}, confirm that the two-dot structure is successfully resolved in parameter regimes with favorable SBR and SNR, and progressively degrades as conditions worsen — consistent with the quantitative predictions of Sec.~4.

The object intensity distributions shown in Fig.~\ref{fig:fig2}(a,b) were mostly intended to reproduce, respectively, what was done in the first IDI experiment~\cite{Trost_PhysRevLett.130.173201} and the first high-order correlation imaging experiment~\cite{Schneider_Nat_phys}, which is restricted to retrieve the finite number of spatial frequencies of a discrete object distribution. The example shown in Fig.~\ref{fig:fig2}(c) is, to our knowledge, the first implementation of IDI on a compact non-periodic object, a letter $\lambda$ in this case. 

\section{Assessing imaging capabilities of IDI}

The versatility of this PTLS optical setup in terms of brightness and degree of coherence can be exploited to benchmark the analytical and numerical models of IDI that have been developed over the recent years~\cite{Trost_2020,Lohse:iv5016}. Of special importance are the scalings of the visibility for different source brightness (in terms of average number of photons per shot per pixel $\mu$) and number of temporal modes accumulated in each frame. For a PTLS, this can be defined as~\cite{goodman1985statistical} 
\begin{equation}
    \mathcal{M}=\left[ \frac{1}{T}\int_{-T}^{T} \left( 1-\left|\frac{\tau}{T}\right| \right) |\gamma(\tau)|^2 d\tau \right]^{-1}
\end{equation}
where $T$ is the camera acquisition time and $\gamma(\tau)$ is the complex degree of coherence, for which we use the fitted function shown in Fig.~\ref{fig:fig1}(b). Note that, when $T\gg\tau_c$, we can approximate $\mathcal{M}\approx \frac{T}{\tau_c}$~\cite{Trost_2020}, i.e. the ratio between the duration of the acquisition  of each frame and the temporal coherence of the light source. 
For simplicity, we are going to restrict our study to the object intensity distribution used for the first experimental realisation of IDI in an FEL facility~\cite{Trost_PhysRevLett.130.173201}, two spatially separated illumination areas as shown in Fig.~\ref{fig:fig2}(a).

In Fig.~\ref{fig:fig3}(a) we plot the outlines of the $g^{(2)}-1$ maps (perpendicular to the fringes) for different number of temporal modes $\mathcal{M}$ accumulated in each acquisition. Experimentally, the number of temporal modes is controlled via the diffuser rotation speed. Other parameters such as detector exposure time (1 ms), average number of photons per pixel per frame $\mu \approx 0.1$ and total number of frames (100) are kept constant. As expected, the contrast of the fringes decreases as $\mathcal{M}$ increases. 

\begin{figure}[h]
    \centering
    \includegraphics[width=0.99\textwidth]{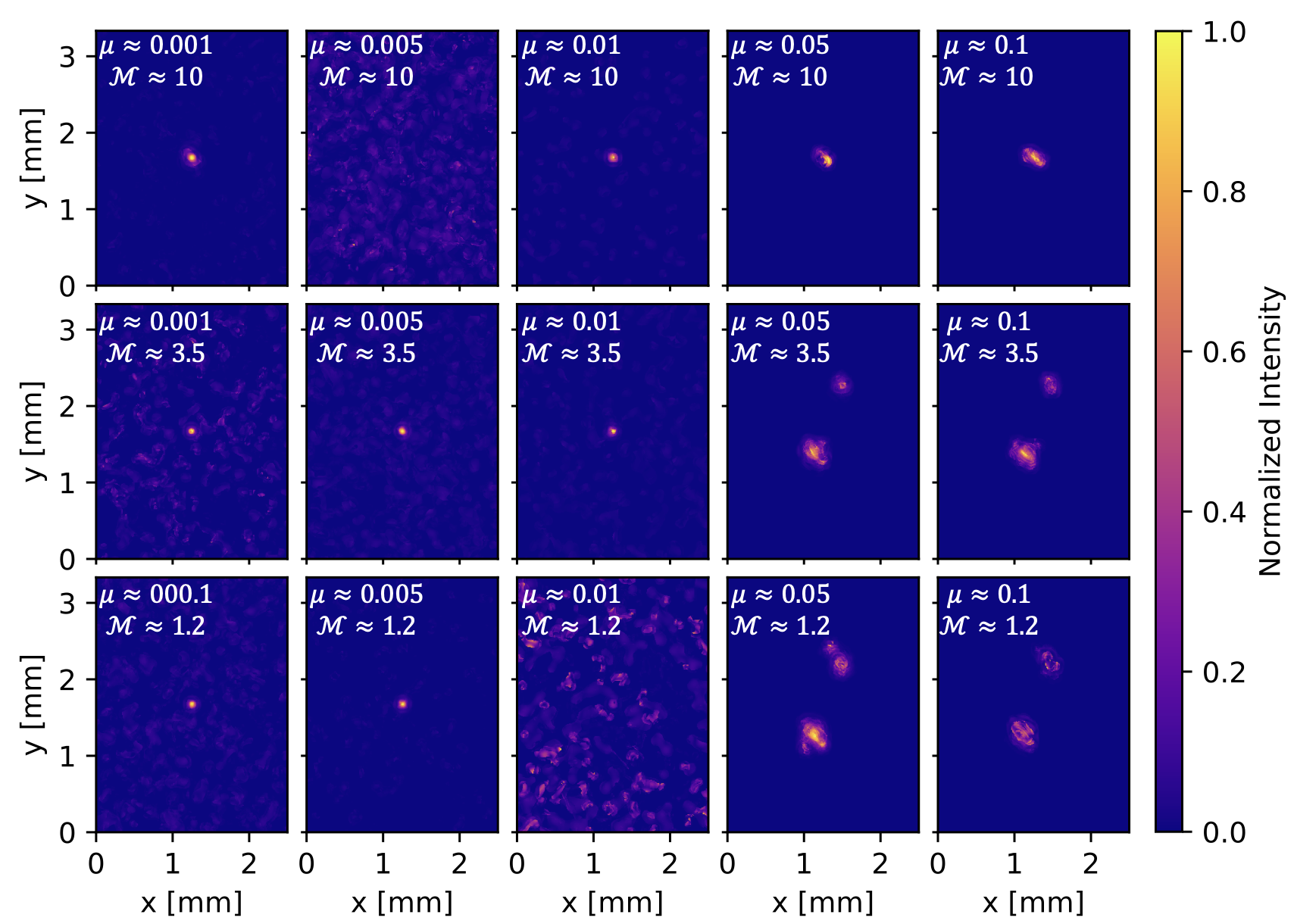}
    \caption{Phase retrieval reconstructions of the two-dot object for different values of average photon flux $\mu$ and number of temporal modes $\mathcal{M}$, spanning the experimental parameter range explored in the main manuscript. Phase retrieval reconstructions of the two-dot object for different values of average photon flux µ and number of temporal modes M, spanning the full experimental parameter range explored in this work. The algorithm is initialised using an autocorrelation-based support estimate. The main reconstruction stage employs the Hybrid Input-Output (HIO) algorithm over 15,000 iterations, with adaptive support updates applied at regular intervals using relative thresholding to progressively refine the support constraint. A final Error Reduction (ER) refinement stage is used to converge toward a consistent real-space density.}
    \label{fig:fig4}
\end{figure}

\begin{figure}[h]
    \centering
    \includegraphics[width=0.99\textwidth]{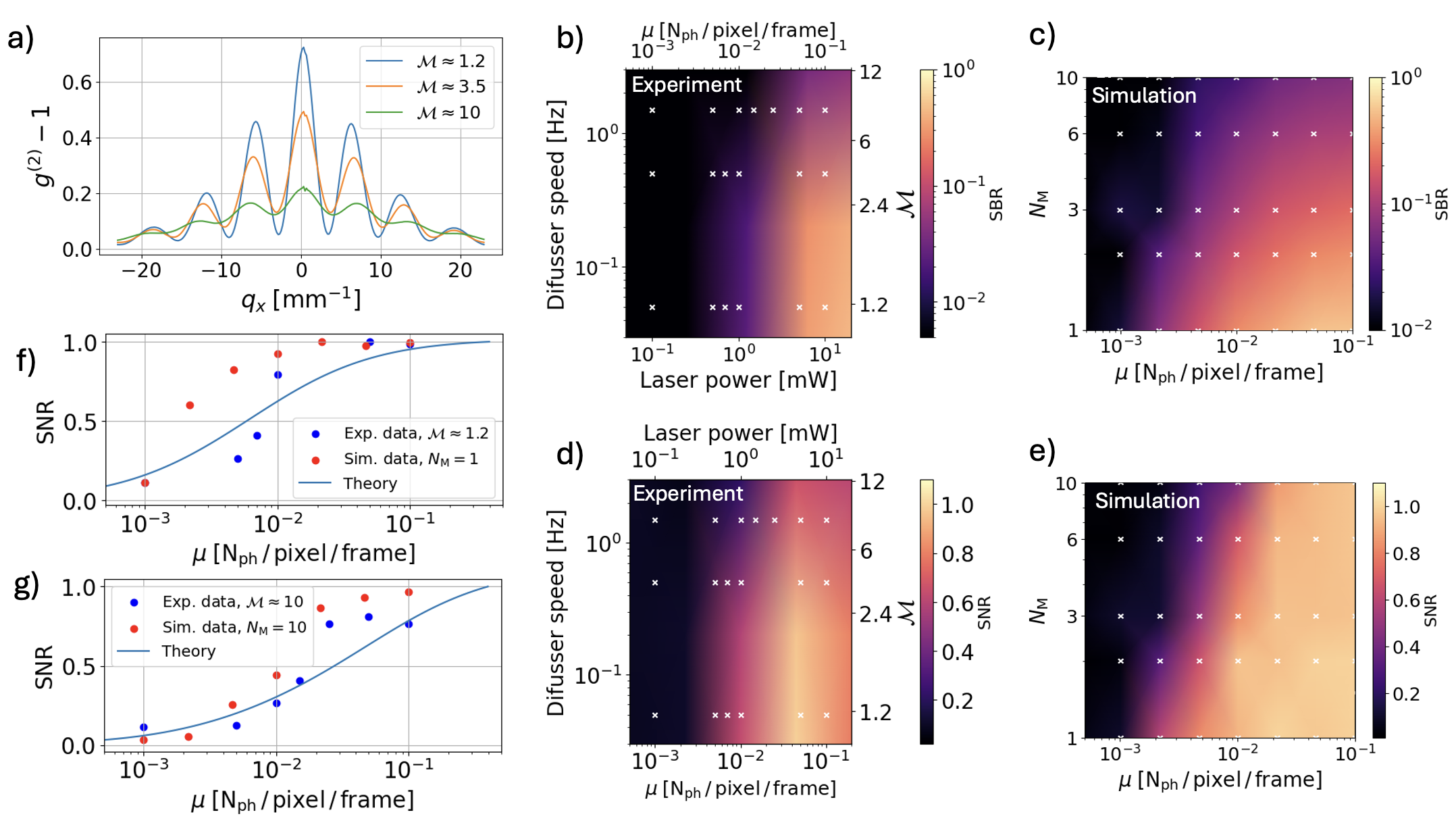}
    \caption{(a) Second order fringes produced by a double Gaussian impinging onto the ground-glass diffuser for different degrees of temporal coherence, i.e. accumulated temporal modes. (b) Measured SBR (see text for details) for different $\mu$ and $\mathcal{M}$ $(N_{\rm M}$). The actual measurements were performed at the white crosses, and the colormap shows interpolated results. (c) Equivalent SBR extracted from simulations. (d) Measured SNR. (e) Equivalent SNR extracted from simulations. (f) SNR as a function of $\mu$ for $\mathcal{M}\approx 1$ as measured from experimental data (blue dots), as extracted from simulations (red dots), and given by analytical model (blue line). (g) Same as (f) but for $\mathcal{M}\approx 10$.}
    \label{fig:fig3}
\end{figure}

Since the signal cannot be readily separated from background and noise, we have followed a similar approach as in~\cite{Trost_2020} for dense objects to quantify the level of visibility and noise under different experimental conditions. We have fitted the $g^{(2)}$ outlines (e.g. Fig.~\ref{fig:fig3}(a)) to the ground truth fringe pattern~\cite{fit_formula} multiplied by an amplitude ($S$) plus a constant background ($B$), i.e. $g^{(2)}=B+S|g^{(1)}(q)|^2$,
with $S$ and $B$ as free parameters. We the define the Signal-to-Background ratio (SBR) as the ratio between S and B, and the Signal-to-Noise ratio (SNR) as the signal divided by the standard deviation of the error, i.e. the difference between the best-fit and the experimental data. We have computed these SBR and SNR for different average number of photons per pixel per frame $\mu$ and different number of accumulated temporal modes. To account for the different $g^{(2)}$ statistics at $\mathbf{q}=0$, this value is excluded from the fits. The results are shown in Fig.~\ref{fig:fig3} (b) and (d). Using 100 frames for the calculation of each $g^{(2)}(\vec{q})$, we measured an maximum SBR of $\approx 0.7$ for a large enough brightness ($\mu\gtrsim 0.1$) and for $\mathcal{M}\sim 1 $ within our parameter range. For the same number of total frames, either increasing the detected number of temporal modes or decreasing the source brightness leads to a decrease of both SBR and SNR. These trends are directly reflected in the phase retrieval reconstructions shown in Fig.~\ref{fig:fig4}: successful resolution of the two-dot structure is achieved in the high-SBR, high-SNR corner of the parameter space ($\mu \gtrsim 0.01$, $\mathcal{M} \lesssim 3.5$) and progressively degrades as either $\mu$ decreases or $\mathcal{M}$ increases. In addition to the SNR and SBR, the Z-score of these $g^{(2)}(\vec{q})$ is displayed in the supplemental document in Fig. S1.

Furthermore, we have used the previously explained simulation framework to compare the experimental measurements of SBR and SNR with numerical simulations performed with the same range of physical parameters. 
The results of these simulations are shown in Fig.~\ref{fig:fig3} (c) and (e). 
Note that in these simulations only an integer number of temporal modes $N_{\rm M}$ can be simulated. Overall, a good agreement is found between these simulations and the experimental measurements. Similar values for the optimal SBR of $\approx 0.7$ and SNR of $\approx 1$ are observed in these simulations at the same optimal conditions within our parameter range, as well as similar trends are also retrieved. Horizontal outlines of these colourmaps for $\mathcal{M}\sim 1$ and $\mathcal{M}\sim 10$, with both experimental and simulation results, are also shown in Fig.~\ref{fig:fig3} (f) and (g). These panels also include the theoretical scaling of the SNR  as a function of $\mu$~\cite{Trost_2020}:
\begin{equation}
    \mathrm{SNR \propto \frac{\mu^2}{\mathcal{M}\sqrt{\frac{1+4\mathcal{M}}{\mathcal{M}^2}\mu^4+2\frac{1+2\mathcal{M}}{\mathcal{M}}\mu^3+\mu^2}}}\,.
\end{equation}
In order to facilitate the comparison between the experimental, simulated and analytical trends, the SNR values plotted in panels (f) and (g) are normalized by the maximum SNR measured. The overall trends observed both in simulations and experimental data seem to follow the theoretical scaling, both with respect to $\mu$ and $\mathcal{M}$.
However, we find that in both cases the theoretical trend tends to underestimate the rate of decrease of SNR when $\mu$ decreases, which could be related to the limited range of applicability of the theoretical model. Moreover, for both $\mathcal{M}\sim 1$ and $\mathcal{M}\sim 10$, the experimentally measured SNR is lower than the simulated value. We have identified two potential sources that could originate this discrepancy: the modeling $\mathcal{M}$ via the accumulation of a discrete number of randomisations $N_{\rm M}$, and contributions from a non-fully characterised (and thus subtracted) noise contributions from some elements of the experimental setup (CMOS detector, laser), which may also account for the non-monotonic dependence of the experimental SNR on $\mu$ observed in Fig. 3(d). Concerning the discrepancy with the theoretical SNR, it should be noted that a similar discrepancy between this scaling and simulation of non-crystalline objects was observed in the aforementioned theoretical study, and was associated to the strong dependence of the SNR on the object complexity.

The parametric study presented here provides direct guidance for the design of future XFEL-based IDI experiments targeting more complex objects than those demonstrated so far \cite{Trost_PhysRevLett.130.173201,radloff_2025_arxiv}. Our results confirm that operating with $\mathcal{M}\lesssim$ a few temporal modes — achievable when the FEL pulse duration is comparable to or shorter than the fluorescence coherence time $\tau_c$ of the target material — is essential for reaching the SBR and SNR levels at which reconstruction succeeds (Fig.~\ref{fig:fig4}). The simulation tool made available alongside this work~\cite{repo_IDI_2024} allows direct mapping of planned XFEL conditions onto the $(\mu, \mathcal{M})$ parameter space characterised here, enabling systematic feasibility assessment prior to beamtime.

\section{Discussion and Conclusions}

In this article, we have presented an original table-top optical setup that allows to validate of Incoherent Diffraction Imaging approach using a Pseudo-Thermal Light Source. This setup has enabled us to experimentally explore IDI over a wide range of physically relevant parameters such as the degree of coherence of the scattered light and the brightness of the light source. Whereas the first experimental realisations of IDI were carried out at FEL facilities, at the wavelength range (X-rays) where this technique holds the promise to unlock high resolution diffraction-based imaging with partially coherent light sources, the restricted access to these facilities limits the amount of relevant experimental data that can be currently produced at this first testing stage of IDI.  To address this challenge, we constructed a scaled experiment using a visible laser, rotating ground glass and spatial light modulator to simulate x-ray incoherent diffraction. This provides an accessible and versatile mean of studying the IDI technique before transitioning to large-scale X-ray facilities. This table-top setup allows us to further explore the key parameters of the IDI approach at scale, offering experimental evidence of the scalings of the Signal-to-Background Ratio and Signal-to-Noise Ratio for different physically relevant conditions, namely low number of photons per pixel per frame, and large number of temporal modes ($\sim 10$) detected at each acquisition. The measured visibility and trends are found to be in reasonably good agreement with existing numerical and analytical methods used to model IDI experiments. Full image reconstruction from the experimentally retrieved $g^{(2)}$ maps, particularly for extended non-periodic objects, remains an exciting direction for future work. 
 
The proposed setup can experimentally benchmark the IDI methodology and thus efficiently address the potential of IDI to realize nanometer scale diffraction-based imaging using partially coherent light sources, before transitioning to fluorescence-based experiments at X-ray facilities.


\noindent\textbf{Funding.} This work was supported by the European Innovation Council (https://ror.org/05cx8cy07) 	under the agreement number 101047223 and by the
Fundação para a Ciência e Tecnologia (https://ror.org/00snfqn58) 	under the project number 2023.14906.PEX\par\medskip

\noindent\textbf{Data availability.} Data underlying the results presented in this paper are available in Ref. \cite{repo_IDI_2024}\par\medskip

\noindent\textbf{Supplemental document.} See Supplement 1 for supporting content.\par\medskip

\bibliography{sample}

\begin{thebibliography}{10}
\bibitem{fit_formula}
$I(\theta)\propto \cos^2(\frac{\pi
  d}{\lambda}\sin\theta)\times\mathrm{sinc}^2(\frac{\pi
  r}{\lambda}\sin\theta)$, $d\approx 1\rm\, mm$ is the distance between the two
  Gaussian light sources, $r\approx 100\rm\,\rm \mu m$ is the root-mean-square
  size of each Gaussian light source and $\theta$ is the observation angle.
\end{thebibliography}

\end{document}